\documentclass[aps,prb,showpacs,twocolumn,byrevtex,superscriptaddress]{revtex4-1}
\usepackage{amsmath}
\usepackage{graphicx}
\usepackage{bm}
\usepackage{multirow}
\usepackage{dcolumn}

\newcommand{\icm}{\ensuremath{\textrm{cm}^{-1}}}
\newcommand{\BFCA}{Ba(Fe,Co)$_2$As$_2$}
\newcommand{\BFCAD}{Ba(Fe$_{0.92}$Co$_{0.08}$)$_2$As$_2$}
\newcommand{\BFCAx}{Ba(Fe$_{1-x}$Co$_x$)$_2$As$_2$}

\newcolumntype{.}{D{.}{.}{-1}}
\begin{document}

\bibliographystyle{apsrev}

\title{Optical signature of sub-gap absorption in the superconducting state of \BFCA}

\author{R. P. S. M. Lobo}
\affiliation{LPEM, CNRS, UPMC, ESPCI-ParisTech, 10 rue Vauquelin, F-75231 Paris Cedex 5, France}

\author{Y. M. Dai}
\affiliation{LPEM, CNRS, UPMC, ESPCI-ParisTech, 10 rue Vauquelin, F-75231 Paris Cedex 5, France}
\affiliation{National Laboratory for Superconductivity, Beijing National Laboratory for Condensed Matter Physics, Institute of Physics, Chinese Academy of Sciences, Beijing 100190, China}

\author{U. Nagel}
\affiliation{National Institute of Chemical Physics and Biophysics, Akadeemia tee 23, 12618 Tallinn, Estonia}

\author{T. R\~o\~om}
\affiliation{National Institute of Chemical Physics and Biophysics, Akadeemia tee 23, 12618 Tallinn, Estonia}

\author{J. P. Carbotte}
\affiliation{Department of Physics and Astronomy, McMaster University, Hamilton, ON L8S 4M1, Canada}

\author{T. Timusk}
\affiliation{Department of Physics and Astronomy, McMaster University, Hamilton, ON L8S 4M1, Canada}

\author{A. Forget}
\affiliation{CEA, IRAMIS, SPEC, 91191 Gif sur Yvette, France}

\author{D. Colson}
\affiliation{CEA, IRAMIS, SPEC, 91191 Gif sur Yvette, France}

\date{\today}
\begin{abstract}
The optical conductivity of \BFCAD\ shows a clear signature of the superconducting gap, but a simple $s$-wave description fails in accounting for the low frequency response. This task is achieved by introducing an extra Drude peak in the superconducting state representing sub-gap absorption, other than thermally broken pairs. This extra peak and the coexisting $s$-wave response respect the total sum rule indicating a common origin for the carriers. We discuss the possible origins for this absorption as (i) quasiparticles due to pair-breaking from interband impurity scattering in a two band $s_{\pm}$ gap symmetry model, which includes (ii) the possible existence of impurity levels within an isotropic gap model; or (iii) an indication that one of the bands is highly anisotropic.
\end{abstract}

\maketitle

%
%
Superconductivity in oxy-pnictides\cite{Kamihara2008} shows a myriad of unconventional properties. As the electron-phonon coupling in these materials is too small to account for the high observed $T_c$ values,\cite{Boeri2008} the presence of a spin-density wave order with $(\pi,\pi)$ momentum fostered the scenario for a spin fluctuation pairing mechanism.\cite{Kuroki2008} Another distinguishing feature of pnictide superconductors is the presence of multiple bands crossing the Fermi energy.\cite{Singh2008} In a multiband superconductor it is natural to expect several superconducting gaps. The earliest observation of two superconducting gaps in the same material was done in Nb doped SrTiO$_3$ which has a sub-Kelvin $T_c$.\cite{Binnig1980} The presence of two superconducting gaps in MgB$_2$,\cite{Giubileo2001} with $T_c \sim 40$~K, made multiband superconductivity accessible for most measurement techniques. \BFCA\ seems to be another example of a high-$T_c$ superconductor with multiple gaps. The unconventional pairing mechanism and the multiband character of \BFCA\ open several possibilities for the gap symmetry with interesting consequences for the optical conductivity. 

Several groups have measured the optical conductivity of \BFCA\ and other pnictides without finding a clear  multigap signature.\cite{Li2008,vanHeuman2009,Kim2009,Gorshunov2010,Fischer2010} Far-infrared measurements by  \textcite{Gorshunov2010} show that, although a gap seems to be present in the optical conductivity, a single $s$-wave Mattis-Bardeen description fails at low frequencies. The data show a residual conductivity that is much higher than that produced by thermally broken pairs. \textcite{vanHeuman2009} showed that the optical conductivity can be described by the superposition of two $s$-wave gaps. However, in their analysis the smaller gap energy falls at the lower end of their measured spectral range. As a result, their fits are dominated mostly by the high-frequency (above the gap) Mattis-Bardeen response. The \textcite{Kim2009} description of the far-infrared optical conductivity, obtained by ellipsometry, takes into account the presence of three superconducting gaps, having values compatible with other techniques. Their overall 5 K spectrum is well described, but does not show any clear signatures at the gap values. 

Band structure calculations suggest that the gap response in \BFCA\ has $s_{\pm}$ symmetry.\cite{Mazin2008} The Fermi surface is composed of different pockets: a hole pocket around the $\Gamma$ point with an $s$ gap having a sign opposite to the gap of the electron pockets at $(\pi,\pi)$. \textcite{Hanaguri2010} have obtained strong evidence for such a sign change in Fe(Se,Te) from the observed magnetic field response of the quasiparticle interference patterns seen in scanning tunneling microscopy. Non-magnetic interband impurity scattering is pair-breaking in $s_{\pm}$ symmetry the same way that Abrikosov-Gorkov (AG)\cite{Abrikosov1961} magnetic impurities are in the isotropic $s$-wave case. The strong scattering limit of AG can give separate impurity levels in the gap\cite{Shiba1968,*Shiba1973,Rusinov1969,*Rusinov1969b,Schachinger1984} which broaden into a band at larger values of doping. \textcite{Vorontsov2009} have noted that this leads to a $T^2$ low temperature penetration depth in agreement with some experiments. Such pair breaking effects provide absorption within the gap in optics.\cite{Nicol1992} Alternatively, as discussed by \textcite{Chubukov2009}, the electron band can be very anisotropic and possibly even have nodes on the Fermi surface. If so, these nodes could be lifted through impurity scattering.\cite{Mishra2009,Carbotte2010} Indeed, such an effect has been observed in the $B_{2g}$ Raman spectra of \BFCAx.\cite{Muschler2009}  

As no clear-cut picture emerges from a multigap fitting of the optical conductivity, here we take a model independent approach to the analysis of the data. We show that the optical conductivity of \BFCAD\ can be described by the superposition of an $s$-wave gap and a Drude term representative of sub-gap absorption in the superconducting state. We discuss the extra optical conductivity in the framework of broken pairs, gap anisotropy or impurity levels in the gap.

%
%
We measured the near normal (10$^\circ$) incidence reflectivity of a \BFCAD\ single crystal on a cleaved $ab$ plane. The sample was grown by a self-flux method\cite{Rullier2009} and showed $T_c = 22.5$~K. The measured surface was $3 \times 3 \textrm{ mm}^2$. The absolute reflectivity was measured in Paris on Bruker IFS113v and IFS66v spectrometers. Data from 20~\icm\ to 8000~\icm\ were collected at several temperatures down to 4~K inside an ARS Helitran cryostat. In order to obtain the absolute reflectivity of the sample, we used an \textit{in situ} gold overfilling technique.\cite{Homes1993} With this technique, we can achieve an absolute accuracy in the reflectivity better than 1\%, and the relative error between different temperatures is of the order of 0.2\%. In order to use Kramers-Kronig analysis to obtain the optical conductivity, we measured the room temperature reflectivity up to 55~000~\icm\ and appended the data to all other temperatures. We completed the low frequency absolute reflectivity data using either a Hagen-Rubens or a two-fluid extrapolation. At high frequencies we used a constant reflectivity up to 200~000~\icm\ and terminated the data with a $\omega^{-4}$ free electron response. Different choices of low and high frequency extrapolations did not change the optical conductivity more than 1\% in the [25--300]~\icm\ range. The very low frequency superconducting-to-normal reflectivity ratios ($R_S/R_N$) were measured in Tallinn down to 5~\icm\ utilizing a SPS200 (Sciencetech, Inc.) polarizing Martin-Puplett interferometer. This setup is  described in Ref.~\onlinecite{Nagel2008}, and it probes the conductivity in the $ab$ plane without  contaminations from the c-axis conductivity.

\begin{figure}[htb]
  \includegraphics[width=8cm]{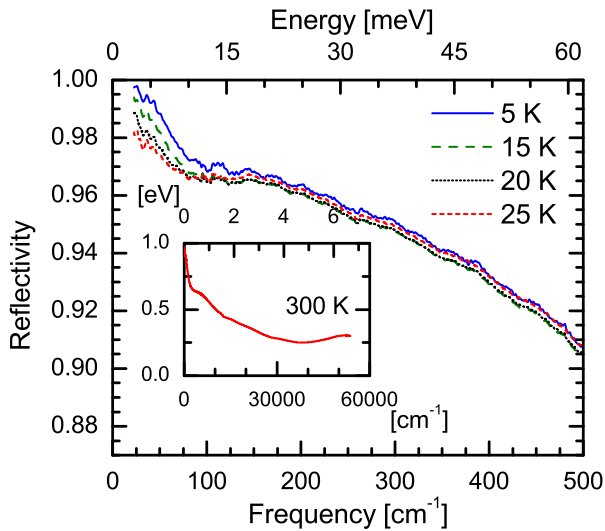} 
  \caption{(color online) Reflectivity of \BFCA\ at low temperatures above and below $T_c = 22.5$~K. The inset shows the 300~K reflectivity in the full measured spectral range.}
  \label{fig_R}
\end{figure}

%
%
Figure~\ref{fig_R} shows the far-infrared absolute reflectivity at low temperatures, above and below $T_c$. Upon crossing the superconducting transition, the reflectivity increases below 100~\icm\ but does not reach a flat unity response expected for a fully $s$-wave gapped superconductor. The inset of this figure shows the reflectivity at 300~K measured in the full spectral range. 
\begin{figure}[htb]
  \includegraphics[width=8cm]{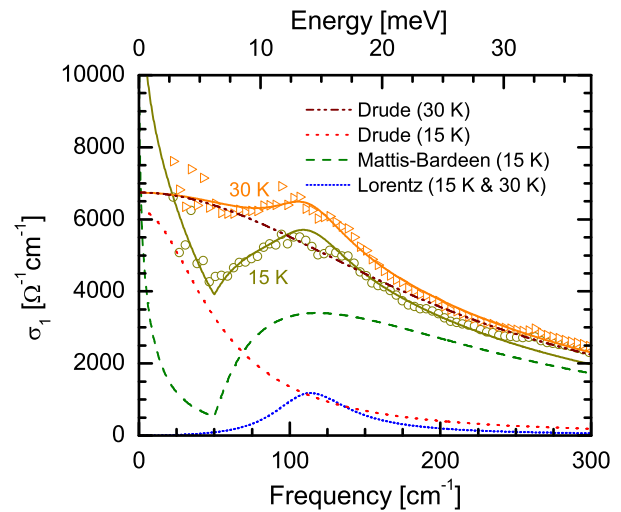} 
  \caption{(color online) The open circles are the real part of the optical conductivity at 15 K and the open triangles at 30 K. The solid lines going through these points are fits to the data using Eq.~\ref{eqmodel}. In the normal state the fit is composed of a Drude term ($\Omega_p = 9250~\icm$; $\tau^{-1} = 210~\icm$) and a Lorentz peak ($\Omega = 114~\icm$; $S = 2071~\icm$; $\gamma = 61~\icm$). In the superconducting state, the same Lorentz peak is kept and the Drude term has its spectral weight divided up between a Mattis-Bardeen $s$-wave gap ($\Omega_p^N = 8210~\icm$; $\tau_N^{-1} = 200~\icm$; $2 \Delta = 50~\icm$) and a residual Drude peak ($\Omega_p = 4450~\icm$; $\tau^{-1} = 53~\icm$).}
  \label{fig_decomp}
\end{figure}
The open symbols in Fig.~\ref{fig_decomp} are the real part of the optical conductivity ($\sigma_1$) determined at 15~K and 30~K. The solid lines are a multicomponent fit for $\sigma_1$, written as:
%
\begin{multline}
  \sigma_1(\omega)=\frac{2\pi}{Z_0} 
    \left[ 
      \frac{\Omega^2_p}{\tau}\frac{1}{\omega^2 + \tau^{-2}} + 
      \frac{\gamma \omega^2 S^2}{\left(\Omega^2-\omega^2\right)^2 + \gamma^2 \omega^2}
    \right] + \\
    \sigma_1^S(\omega, \Omega_p^{N}, \tau_N, \Delta, T),
\label{eqmodel}
\end{multline}
%
where $Z_0$ is the vacuum impedance. The first term in Eq.~\ref{eqmodel} corresponds to a Drude response of unpaired carriers; the second term to a Lorentz oscillator describing a finite frequency resonance; and the last term ($\sigma_1^S$), which exists only below $T_c$, is the optical conductivity for an $s$-wave superconductor. The Drude response is characterized by a plasma frequency ($\Omega_p$) and a scattering rate ($\tau^{-1}$). The Lorentz oscillator is defined by a resonance frequency ($\Omega$) a line width ($\gamma$) and a plasma frequency ($S$). We took $\sigma_1^S$ in the convenient form proposed by \textcite{Zimmermann1991} Besides the temperature ($T$) and the superconducting gap ($\Delta$), it also depends on the Drude weight ($\Omega_p^{N}$) and the scattering rate ($\tau_N^{-1}$) that the carriers would have, had the system been driven normal below $T_c$.

We fitted the 30~K data assuming a single Drude response and a Lorentz oscillator. The Drude term characterizes the free carriers. The origin of the Lorentz peak is not well established. It could be the response of localized carriers induced by disorder but, as it has been seen in an independent \BFCA\ measurement,\cite{vanHeuman2009} it is likely an intrinsic excitation such as low energy interband transitions. In any case, its spectral weight is small and the parameters used in this Lorentz peak at 30 K were kept fixed at all other temperatures. 

In a conventional BCS superconductor, one would replace the normal state Drude term by a Mattis-Bardeen response alone. However, to describe the data at 15~K, we must keep an independent Drude peak in the superconducting state. The fit to the 15~K data is then composed of the same Lorentzian found at 30~K together with a Mattis-Bardeen \textit{and} a Drude peak. The effect observed by \textcite{Gorshunov2010}, namely that the measured low frequency $\sigma_1$ is higher than the thermally broken pairs in a Mattis-Bardeen term, is clearly shown in Fig.~\ref{fig_decomp}. Adding the Drude response in the superconducting phase (red dotted line) to the Mattis-Bardeen component (green dashed line), leads to a proper description of the data below 50~\icm. Note that the measured $\sigma_1$ has a low frequency upturn with a width ($\sim 50~\icm$) that is much broader than the width ($\sim 10~\icm$) of the thermally broken pairs from the BCS contribution. Hence, the Mattis-Bardeen description fails below the gap and the low frequency $\sigma_1$ requires the additional Drude peak in the superconducting state. Before proceeding, we must remark that in order to minimize the number of parameters, we utilize a single BCS gap in our description. Introducing extra BCS gaps produces equivalent fits which cannot account for the low frequency absorption either. The presence of an extra Drude term, having the same weight and width as the one used for the single gap fits, is still necessary.

\begin{figure}[htb]
  \includegraphics[width=8cm]{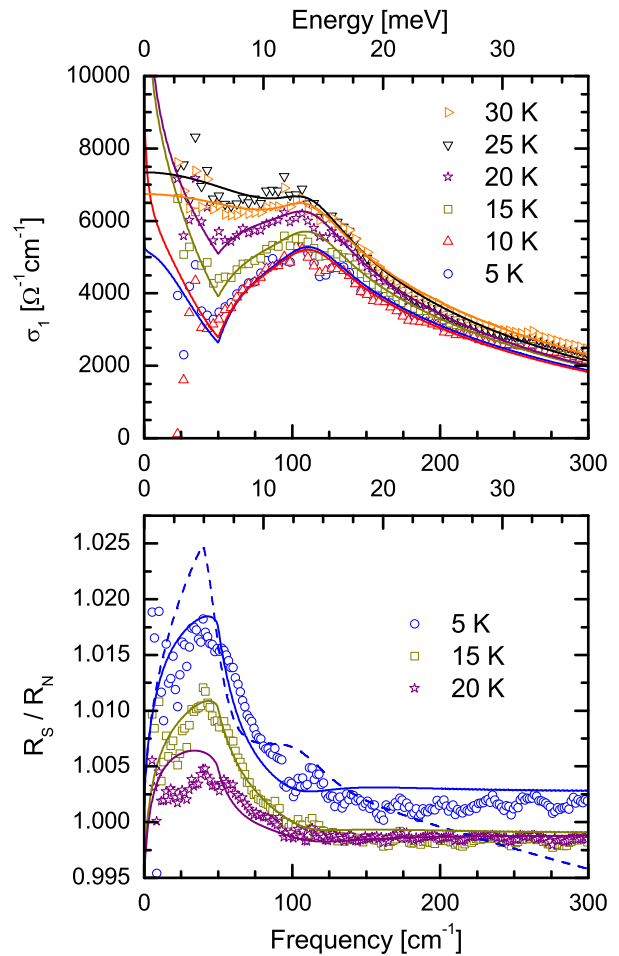} 
  \caption{(color online) The top panel shows the measured $\sigma_1$ (symbols) at various temperatures and corresponding fits (solid lines) using Eq.~\ref{eqmodel}. The symbols in the bottom panel are the measured superconducting-to-normal reflectivity ratios ($R_S/R_N$) and the simulations obtained by using the parameters that fit the optical conductivity. $R_N$ is taken at 25~K. The blue dashed line is the best fit assuming that the superconducting state has no unpaired quasiparticles due to scattering pair breaking (no Drude term).}
  \label{fig_fits}
\end{figure}
The top panel in Fig.~\ref{fig_fits} shows the above model applied to the measured temperatures, up to 30~K. We first fitted the 30~K data. For lower temperatures we only allowed three parameters to vary: the plasma frequency and the scattering rate of the Drude peak as well as the spectral weight $[(\Omega_p^N)^2]$ of the Mattis-Bardeen contribution. For the latter term, we used as input parameters the measured temperature and a superconducting gap of $2\Delta = 50~\icm$, an average value in agreement with other measurements in the same system.\cite{Kim2009,vanHeuman2009,Fischer2010} The scattering rate $\tau_N^{-1}$ was also taken as a fixed parameter and its value was set to 200~\icm\ --- the average of the scattering rates obtained at 25~K and 30~K. 

The open symbols in the bottom panel of Fig.~\ref{fig_fits} show the reflectivity ratios $R_S/R_N$, where $R_N$ is taken at 25~K. The solid lines are calculations utilizing the same parameters as those in the upper panel. The dashed line is an attempt to make a fit to the 5~K data with an $s$-wave gap and no residual Drude peak. Here, again, the presence of an extra Drude term in the superconducting state is paramount to the description of the data.

This approach is clearly a convenient way to parametrize the optical response of \BFCA. But does it represent a more fundamental physical interaction? The assumption in introducing a residual Drude term in the superconducting state was that extra states, other than thermally broken pairs, exist below $T_c$. We remark that the Drude profile obtained for the sub-gap absorption is inconsistent with separate finite energy impurity levels in the gap. So, we will focus our discussion in the pair breaking and the gap anisotropy scenarios. 

\begin{figure}[htb]
  \includegraphics[width=8cm]{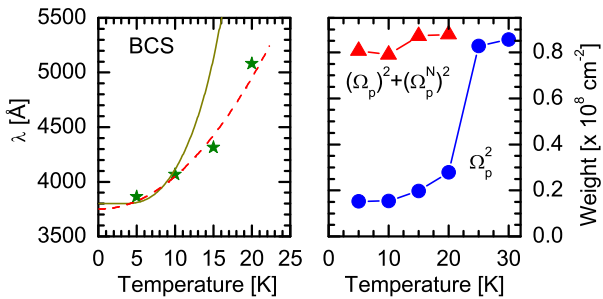} 
  \caption{(color online) The symbols in the left panel depict the temperature evolution of the penetration depth calculated from the imaginary part of the optical conductivity. The solid line is a single gap BCS calculation with $2\Delta = 50\icm$. The dashed line is a quadratic $\left(\Delta\lambda= 3.2 \, T^2\right)$ behavior. The right panel shows the application of the $f$-sum rule to \BFCA. The solid blue circles are the spectral weight of the Drude term in both normal and superconducting states. The solid triangles represent the total weight in the superconducting state, including the superfluid stiffness.}
  \label{fig_sumrule}
\end{figure}

The solid stars in the left panel of Fig.~\ref{fig_sumrule}, show the thermal dependence of the penetration depth calculated from the imaginary part of the optical conductivity. The BCS behavior, shown as the solid line, cannot describe the data. The dashed line is the $T^2$ behavior compatible with nodes in the gap\cite{Fischer2010} or a multigap system with pair-breaking interband scattering.\cite{Vorontsov2009} To get a better insight into these two pictures we can use the optical conductivity $f$-sum rule:
%
\begin{equation}
  \int_0^\infty \sigma_1(\omega) d\omega = \frac{\pi^2}{Z_0} \Omega_p^2.
	\label{eqsumrule}
\end{equation}
%
Equation~\ref{eqsumrule} implies charge conservation and is independent of external parameters such as the temperature. In a conventional BCS superconductor, the weight of the normal state Drude peak ($\Omega_p^2$) is fully transferred to the Mattis-Bardeen weight $[(\Omega_p^N)^2]$. 

The right panel of Fig.~\ref{fig_sumrule} shows the $f$-sum rule applied to our model. The solid circles are the temperature evolution of the spectral weight of the normal state Drude peak and the superconducting state residual term. At $T_c$ there is a large drop ($\sim75\%$) in the spectral weight of the unpaired carriers, but it does not vanish in the superconducting state. When we add the spectral weight of the Mattis-Bardeen term, we find the line defined by the solid triangles. This figure shows that the total spectral weight is conserved and that the normal state Drude term exactly transfers its weight to the residual unpaired quasiparticles plus the particles participating in the Mattis-Bardeen superconducting response. \textcite{Vorontsov2009} estimated that impurity scattering leads to a $T_c$ of 30--60\% of its clean limit value in \BFCA. \textcite{Nicol1992} calculated $\sigma_1$ for a superconductor in the presence of pair-breaking. From their results, we can calculate that the residual sub-gap spectral weight for the $T_c$ drop estimated in \BFCA\ is in the 3--8\% range. Therefore, the observed 25\% remaining spectral weight is more consistent with a highly anisotropic gap in the electron pocket,\cite{Chubukov2009,Mishra2009,Carbotte2010,Muschler2009,Wu2010} which could have nodes or a very small effective gap. Optical data cannot, on its own, differentiate between these two models for the electron pocket, although recent thermal conductivity data argue for a small gap possibly less than 1 meV.\cite{Reid2010} 

%
%
We measured the far-infrared optical conductivity of \BFCAD\ in the normal and superconducting states. We found a clear signature of the superconducting gap but a plain Mattis-Bardeen $s$-wave approach fails to describe the low frequency optical conductivity. The introduction of an additional Drude peak, which accounts for the additional low energy absorption, reconciles the model and the data. We show that this extra peak and the coexisting $s$-wave term respects the total sum rule indicating a common origin for the carriers in both responses. This extra absorption could be due to gap anisotropy of the electron pocket;\cite{Chubukov2009,Mishra2009,Carbotte2010,Muschler2009} impurity localized levels inside an isotropic gap;\cite{Shiba1968,Shiba1973,Rusinov1969,Rusinov1969b,Schachinger1984} or pair-breaking due to interband impurity scattering in an $s_{\pm}$ symmetric gap.\cite{Nicol1992,Vorontsov2009} When we look into the spectral weight of the residual Drude term, we find that it is closer to the predictions of the anisotropic extended $s$-wave picture.

\section*{Acknowledgment}

We thank F. Rullier-Albenque for comments and discussions. Work in Paris had the support from the ANR Grant No. BLAN07-1-183876 GAPSUPRA. Work in Tallinn was supported by the Estonian Ministry of Education and Research (SF0690029s09) and Estonian Science Foundation (ETF7011, ETF8170).

%
%
\bibliography{biblio}

\end{document}